\documentclass{ws-procs9x6}
\usepackage{graphicx}

\newcommand{\hp}{\hat{\rm \bf n}}
\newcommand{\hO}{{\hat{\rm \bf e}_{z}\,\!}}

\newcommand{\wg}{{\omega_{\rm g}}}

\begin{document}

\title{Pulsars and Gravitational Waves}

\author{K. J. Lee\footnote{Email: kjlee007@gmail.com}, %
R. X. Xu and G. J. Qiao}

\address{School of Physics and State Key Laboratory of Nuclear Physics and
Technology,\\
Peking University, Beijing, 100871, P. R. China}

\begin{abstract}
The relationship between pulsar-like compact stars and gravitational
waves is briefly reviewed. Due to regular spins, pulsars could be
useful tools for us to detect $\sim$nano-Hz low-frequency
gravitational waves by pulsar-timing array technique; besides, they
would also be $\sim$kilo-Hz high-frequency gravitational wave
radiators because of their compactness. The wave strain of an
isolate pulsar depends on the equation state of cold matter at
supra-nuclear densities. Therefore, a real detection of
gravitational wave should be very meaningful in gravity physics,
micro-theory of elementary strong interaction, and astronomy.
\end{abstract}


\bodymatter

\section{Introduction}

Since the human mind first wakened from slumber, it has never ceased
to feel the profound nature of space-time, especially the
time-consciousness, in both philosophy and physics. However, a
physical concrete of space-time is clarified only after Einstein's
first insight: the space-time is a four dimensional continuum and
the rule of the motion is the pure geometrical constrain that free
particles follow the geodesics of the space-time, while the response
of the space-time continuum to the matter is determined by the
Einstein's equation in which a linear function of space-time
curvature is in proportion to the energy-momentum of the matter.

It is worth noting that the nature of space-time is a debating topic
starting earlier than the general relativity and would not be
terminated only by Einstein's pure geometrical arguments. Guided by
different perceptions of space-time philosophies, many gravity
theories are proposed~\cite{Will06}, with different interpretations
of equivalence principle. Even in the general relativity, the
equivalent principle is still a matter of debate.
Therefore, experimental tests of gravity theories, including in
strong field and with fast motion, are critical to differentiate or
falsify the gravity theories. Such experimental environments are
only available in astrophysics, especially related to compact stars
known as white dwarfs, pulsars/neutron stars, and black holes.

Topics of gravitational waves relevant to the pulsar astronomy are
focused on in this review. The binary pulsar tests for gravity
theories are given in \S\ref{sec:binpsr}. Pulsars as tools of
detecting and as sources of gravitational waves are presented in
\S\ref{sec:gwdecpsr} and \S\ref{sec:source}, respectively. Future
prospects are discussed in \S\ref{sec:fur}.

\section{Pulsars, binary pulsars, and tests of gravity theories}
\label{sec:binpsr}

Soon after the discovery, pulsars are identified as a class of fast
rotating rather than pulsating compact objects. It was used to
believe that pulsars are neutron stars composed by hadronic matter,
because of the very limited knowledge in sub-nucleon research at
that time, but this view might not be true\cite{LP04} since the
lowest compact states could be of quark matters with strangeness
rather than that of neutron liquid.
Observationally, there are two main categories of pulsars: the
millisecond pulsars (MSPs) and the normal pulsars. Normal pulsars
have rotational period from a few tens of milliseconds to a few
seconds, while the MSPs' periods range from about 1 millisecond to a
few tens milliseconds \cite{Lorimer05}. Long term timing monitoring
shows that the MSPs are more stable rotators compared to the normal
pulsars, which might due to both observational reasons and pulsar
intrinsic physics. For MSPs, the difference between model-expected
time of arrival (TOA) of radio pulses and observed TOA is usually
less than 10 percent of their periods, and the most stable MSPs can
achieve $\sim 100$ ns level on the time scale of a few
years~\cite{VBCHV09}.

It turns out that most of the MSPs are in binary system. One
particular interesting system is the recently discovered binary
pulsar system, J0737-3039AB, where both of the two stars are radio
pulsars~\cite{LBK04}. The J0737 is also a highly relativistic
celestial system. Binary pulsars with possible pulsar companions are
listed in Table 1, obtained from ATNF pulsar catalog~\cite{MHTH05}.
\begin{table}\tbl{Parameters for possible pulsar-neutron star systems}
    {\begin{tabular}{c c c c c}
\hline  \hline Name & $P$ & $P_{b}$ & $a$ & $e$ \\
    \hline
 J0737-3039A    & 0.022699      & 0.1023      & 1.4150  & 8.778e-02 \\
 J0737-3039B    & 2.773461      & 0.1023      & 1.5161  & 8.778e-02 \\
 J1518+4904     & 0.040935      & 8.6340     & 20.0440  & 2.495e-01 \\
 B1534+12       & 0.037904      & 0.4207      & 3.7295  & 2.737e-01 \\
 J1756-2251     & 0.028462      & 0.3196      & 2.7564  & 1.806e-01 \\
 J1811-1736     & 0.104182     & 18.7792     & 34.7827  & 8.280e-01 \\
 B1820-11       & 0.279829    & 357.7620    & 200.6720  & 7.946e-01 \\
 J1829+2456     & 0.041010      & 1.1760      & 7.2380  & 1.391e-01 \\
 J1906+0746     & 0.144072      & 0.1660      & 1.4202  & 8.530e-02 \\
 B1913+16       & 0.059030      & 0.3230      & 2.3418  & 6.171e-01 \\
 B2127+11C      & 0.030529      & 0.3353      & 2.5185  & 6.814e-01 \\
 \hline \hline
 \end{tabular}}
 \begin{tabnote}{$P$ is the pulsar period in unit of second, $P_{b}$ is the orbit period in unit of days, $a$ is the projected semi major axis in
     unit of light seconds, and $e$ is the orbital eccentricity.}
 \end{tabnote}
 \label{tab:binpsr}
\end{table}

Armed with such a kind of stable celestial clocks (i.e., pulsars)
relativistically orbiting their companions, one can then test
gravity theories in the case of strong gravitational fields, as
illustrated in the classical system of PSR B1913+16~\cite{DT92}. In
the J0737 system, two pulsars orbit each other with a period of 2.5
hours and a very low orbital eccentricity. Up to known, this J0737
system becomes the most relativistic binary system, the details of
which can be found in the review by Kramer and Wex
(2009)~\cite{KW09}.

To test the gravity theories, one must compare the predicted TOA of
a theoretical model with the observation. In this way, one needs
thus the binary motion dynamical models, in which we put the gravity
theory in. One can then calculate the theoretical pulse TOA at the
solar barycenter. Note that, from a pulsar to the barycenter,
various processes set in, including the photon propagation effects
due to the gravitational field of both the binary system and solar
system, the dispersion of pulsar radio signal due to interstellar
medium and solar wind, and so on. One needs also the solar system
ephemeris to convert the pulsar TOA at the radio telescope to the
barycenter. The modeled TOA will then be compared with observed ones
to see if the gravity theories is able to account for the
observation.

In reality, thanks to the phenomenological framework of
post-Keplerian (PK) parameters, with which gravity theories can be
approximated, we can independently measure these PK parameters by
fitting the TOA data. A gravity test is then via checking the
self-consistency of PK parameters for a particular gravity theory.
There are 7 PK parameters which are possibly measurable in the near
future: advance of periastron $\dot\omega$, gravitational redshift
parameter $\gamma$, Shapiro delay parameters $r$ and $s$, orbit
period derivative $\dot{ P_{\rm orb}}$, spin-orbital coupling
induced precession $\Omega_{\rm so,p}$ and relativistic orbit
deformation $\delta_{\rm \theta}$. These PK parameters, except
$\delta_{\rm \theta}$, have been measured in the 0737 system.

All the 7 parameters measured are functions of two unknown
parameters: pulsar masses, $m_a$ and $m_b$. A double neutron star
system is then overdetermined if one detects three or more PK
parameters. It is worth noting that the double pulsar system of
J0737 offers an extra Keplerian constrain, the mass ratio between
two star, $R(m_a, m_b)=m_a/m_b=x_b/x_a$, with $x$ the projected
semimajor axes. Two recent reviews~\cite{KW09,Stairs03} are valuable
in the topic of testing gravity with binary pulsars.

\section{Detecting gravitational waves with pulsars}
\label{sec:gwdecpsr}

Directly detecting gravitational wave (GW) is the Holy Grail of
present experimental researches, not only in gravity physics but
also in astronomy. With the efforts since 1960s~\cite{FZW61}, recent
equipments (e.g., LIGO~\cite{LIGO09Rev}) may finally allow us to
directly detect GWs although there is no confirmed detection now
yet. In this section, we will review the ability of detecting
gravitational waves using {\em pulsar timing array} (PTA). Potential
roles of testing gravity with PTA are also presented here.

GW is actually a perturbation of space-time, fully characterized by
a wave-like metric perturbation. Detecting GW is thus identical to
measure the wave-like metric perturbation\footnote{It should be born
in mind that detecting of GW is not detecting any types of metric
perturbation. GW detection focuses on detecting the oscillatory part
of the metric perturbation with strain $h$ decrease as $r^{-1}$,
such that gravitational wave could carry energy and momentum to the
infinity.} which can be performed by comparing geodesics of two test
objects approaching to and departing from each other. Such
experiments fall into four categories: 1. Tracing the motion of two
free-falling test objects (e.g. LIGO, LISA, GEO, TAMA, and so on),
2. Detecting the deformation of finite extend solid body (e.g. Bar
detector, Sphere detector, and so on), 3.  Measuring the Doppler
shift of electromagnetic signals from distance free-falling objects
(e.g. Doppler tracking of satellite, pulsar timing array, laser
ranging, LISA), 4. Checking the perturbation of a cosmological
system (e.g. cosmic background B mode detection, weak lensing
survey). Among all these possible ways, PTA is one of the promising
techniques to directly detect gravitational waves, being unique to
detect GW at nano-Hertz band~\cite{JHLM05}.

As we have shown, MSPs are very stable celestial clock in the
Galaxy. GWs perturb the background space-time of the Galaxy, such
that pulsar pulse signals get red or blue Doppler shift along the
path from pulsar to earth. It turns out that such GW-induced
frequency shift only involves the metric perturbation at the pulsar
and that at the earth. The GW-induced frequency shift can be
obtained to be~\cite{LJR08}
\begin{equation}\frac{\Delta\omega(t)}{\omega}=\frac{ \hp^{i}\hp^{j} }{2\left
    (1+{\rm \bf n_{g}} \cdot \hp\right )}  \left[
    h_{ij}(t,0)-h_{ij}(t-
    D/c, \rm \bf D)\right],\label{eq:z}\end{equation}
    where $\omega$ is the pulsar angular frequency of spin, $\hp$ is the
    pulsar
    direction, ${\rm \bf D} = D \hp$ with $D$ the distance to the pulsar,
    $\rm \bf n_{\rm g}$ is the GW propagation direction, and $h$ is the
    perturbation of metric. The GW-induced timing residuals in pulsar TOA
    is therefore
    $R=\int \Delta\omega/\omega \,dt$.

Due to the intrinsic noises and possible non-modeled accelerations
of pulsars, it is unlikely that one can use $R(t)$ of a single
pulsar to detect GWs. Nevertheless, magic happens if we correlate
the residuals of $R_j$ and $R_j$ of two pulsars.
From Eq.~(\ref{eq:z}), one can have a correlation of $\langle R_{i}
R_{j}\rangle= C(\theta) \sigma^2$ for two different pulsars in
general relativity, with $\sigma$ the RMS (root-mean-square) of a
single pulsar's residual. Note that the correlation $C(\theta)$ is a
determined function only involving the angular, $\theta$, between
two pulsars~\cite{HD83}, and this correlation $C(\theta)$ certainly
plays a vital role in detecting GW using a array of pulsar timing
data (PTA) since the shape of $C(\theta)$ is uniquely determined by
a gravity theory and there is no other physical processes to make
the pulsar signal correlated for two pulsars widely separated with a
distance of several thousand light years away from each
other\footnote{The imperfectness of terrestrial clock and un-modeled
solar system dynamics may introduce also correlation between
measured $R_i$ of pulsars, however the angular dependence of such
correlations is very different from that $C(\theta)$ presented in
Eq.(2).}. In the general relativity theory of gravity, fortunately,
the correlation $C(\theta)$ has a very simple form of~\cite{HD83,
LJR08}
\begin{equation}
    C(\theta)=\frac{3x\log
    x}{2}-\frac{x}{4}+\frac{1}{2}\left(1+\delta(x)\right),
\end{equation}
where $x=(1-\cos\theta)/2$.

\begin{figure}
\psfig{file=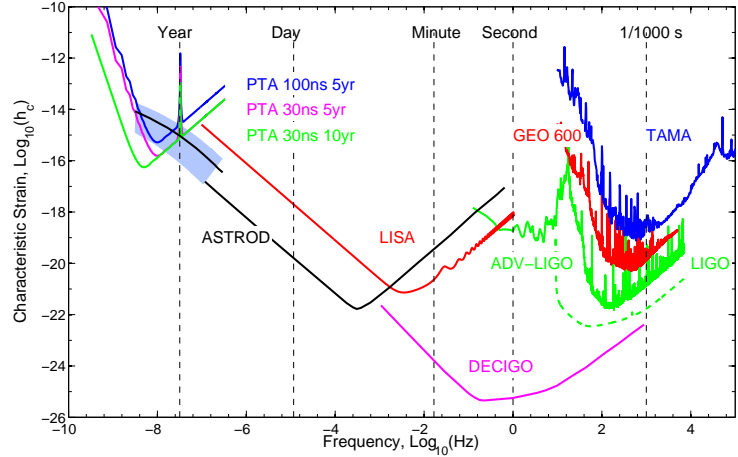,width=4.5in} \caption{A comparative for abilities
of detecting GWs for various GW detectors, where the x-axis and
y-axis are the GW frequency and the characteristic strain of GW
respectively. Most of the labeling are self-explanatory.  The PTA
curves are at in the left-top corner, where the length of data and
pulsar average timing noise level are also labeled.}
\label{fig:sens}
\end{figure}

We may make sense of the $C(\theta)$-curves from simple symmetric
reasons.
If a monochromatic general relativistic GW is propagandizing along
`z-axis' direction (there will be $180^{\circ}$ symmetry and
$90^{\circ}$ anti-symmetry in x-y plane), then correlation
$C(\theta)$ between two pulsars with $0^{\circ}$ or $180^{\circ}$
angular separation are positive, while $C(\theta)$ for $90^{\circ}$
will be negative. This make a U-shaped $C(\theta)$~\cite{HD83}. Note
that $C(180^{\circ})\neq C(0)$ which will be explained later. One
can then measure such multi-pulsar correlation to detect GWs. Jenet
et al. (2005)~\cite{JHLM05} had investigated the statistical
properties of such detection processes. Their results show that
regular timing observations of 40 pulsars each with a timing
accuracy of 100 ns will be able to make a direct detection of the
predicted stochastic background from coalescing black holes within 5
years. We compare the detection abilities for GW detectors in
Fig.~\ref{fig:sens}, for GW background due to coalescing
supermassive binary black holes (BBH).

\begin{figure}
\psfig{file=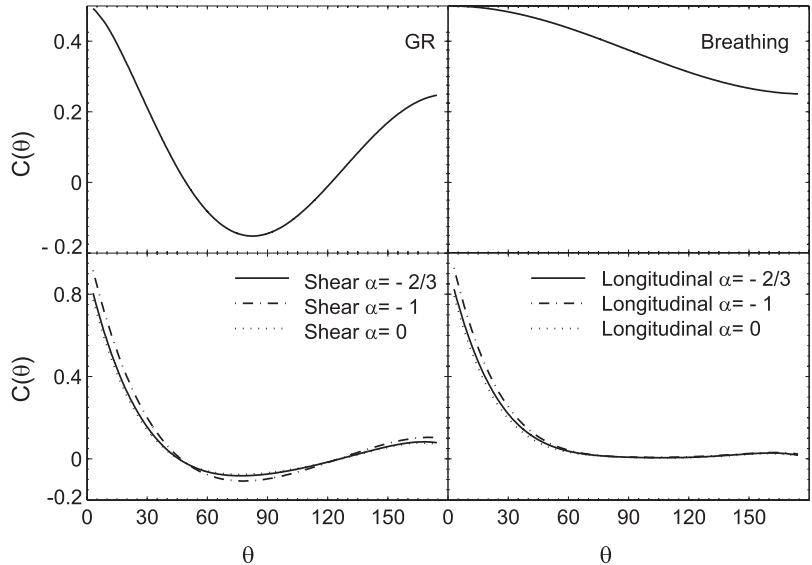,width=4.5in} \caption{The $C(\theta)$ curves for
different kinds of GW polarization, with $\alpha$ denoting the power
index of the GW background.} \label{fig:help}
\end{figure}

From symmetry arguments above, it is clear that the shape of
$C(\theta)$ depends on the polarization of GW (see
Fig.~\ref{fig:help}).  Einstein¡¯s theory of gravity predicts waves
of the distortion of space-time with two degrees of polarization;
alternative theories predict more polarizations, up to a maximum of
six~\cite{ELL73}. Lee et al. (2008)\cite{LJR08} analyzed such
polarization effects and conclude that for biweekly observations
made for five years with rms timing accuracy of 100 ns, detecting
non-Einsteinian modes will require: 60 pulsars in the case of the
longitudinal mode; 60 for the two spin-1 `shear' modes; and 40 for
the spin-0 `breathing' mode. Further more, they showed that one can
test gravity theories by checking GW polarization, i.e., to
discriminate non-Einsteinian modes from Einsteinian modes, we need
40 pulsars for the breathing mode, 100 for the longitudinal mode,
and 500 for the shear mode. These requirement is beyond present
observation technology, but could be easily achieved using SKA or
FAST telescope~\cite{NWZZJG04,SLKMSJN09}.

\begin{figure}
\psfig{file=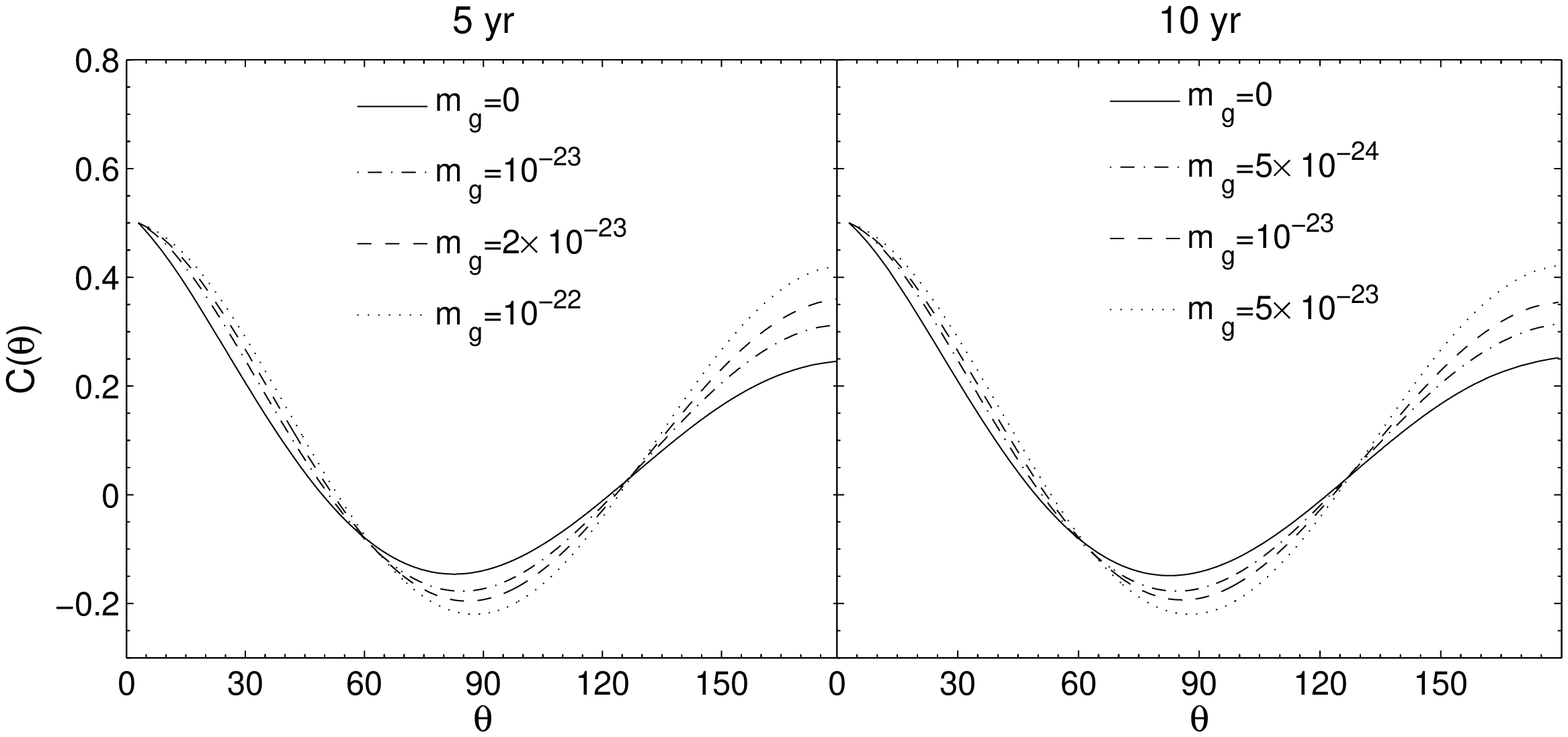,width=4.5in} \caption{The curves of $C(\theta)$
for different graviton masses.} \label{fig:hmass}
\end{figure}

Another interesting topic on detecting GWs using PTA is about the
dispersion relation of GW~\cite{Lee09}, since the function of
$C(\theta)$ and the detection statistics depends also the mass of
graviton. It is found~\cite{Lee09} that $C(180^{\circ})$ increases
to match the value of $C(0)$ as the graviton mass increases (see
Fig.~\ref{fig:hmass} for details). In the case of massless GW
background, we know that the GW has $180^{\circ}$ degree symmetry
due to the polarization property, but why $C(0^{\circ})\neq
C(180^{\circ})$? It turns out that GW propagation breaks up this
$180^\circ$ symmetry by the geometric factor in Eq.~(\ref{eq:z}),
which reads $1+\hO \cdot \hp$ for the massless case. For the case of
a massive GW background, the geometric factor reads
$1+\frac{c}{\wg}{\rm {\bf n}_g} \cdot \hp$, where the graviton mass
reduces the asymmetry. For the limiting case, where the GW frequency
is just at the cut-off frequency, the dispersion relation tells us
that such a GW is not propagative, then the $180^{\circ}$ symmetry
is restored. Therefore, we would expect that the correlation
function are of $180^{\circ}$ symmetry for very massive gravitons.

Lee et al. (2009) further find that it is possible to measure
graviton mass using PTA and one will get 90\% probability to
differentiate between the results for massless graviton and that for
graviton heavier than $3 \times 10^{-22}$ eV, if biweekly
observation of 60 pulsars are performed for 5 years with pulsar RMS
timing accuracy of 100 ns in the future.

As we have shown that PTA can be constructed to measure the
alternative polarization modes of GW and the GW dispersion relation.
These measurements provide tests for gravity theory in the weak
field/high velocity region, which are different from that of the
solar system tests (i.e., the weak field/slow velocity case) and the
binary pulsar tests (i.e., the strong field/slow velocity cases),
because it is not completed to describing GW using post-Newtonian
formalism and the scalar and tensor sectors of gravity theories are
different~\cite{Maggiore08}.

\section{Gravitational wave radiation from pulsars}
\label{sec:source}

Pulsars are not only as tools to detect GWs, but also strong GW
sources because of their compactness and the rapid mass changes.
Indirect evidence for GW from binary pulsars has been discussed in
\S \ref{sec:binpsr}, whereas a direct detection of GW from pulsars
with ground-based facilities should be meaningful. It is recognized
that the GW amplitudes of isolate pulsars depend on the equation of
state (EoS) of cold matter at supra-nuclear density, which is
strongly related to the understanding of QCD (quantum
chromo-dynamics) at low energy scale, still another challenge for
physicists today.
A mixture of quantum (QCD) and gravity (relativity) makes this
project more funny.

We are still not sure about the nature of pulsar-like compact stars
though discovered since 1967. It is conventionally believed that
these compact stars are normal neutron stars composed of hadronic
matter, but one can not rule out the possibility that they are
actually quark stars of quark matter\cite{LP04} (see, e.g., a
review~\cite{xu09}).
Quark stars with strangeness are popularly discussed in literatures,
which are called as strange (quark) stars. The EoS of realistic
quark matter in compact stars, depending non-perturbative QCD, was
supposed to be of Fermi gas or liquid, but could be of classical
solid in order to understand different manifestations of pulsar-like
stars~\cite{xu03,xu09csqcd}.

Besides QCD, that pulsars are quark stars should also be meaningful
in GW physics~\cite{xu06}.
(i). {\em GW being EoS-dependent}. Rotation ($r$) mode instability,
which would result in GW radiation, may occur in fluid quark stars
if the bulk and shear viscosities of quark matter is not
sufficiently high, but no $r$-mode instability occurs in solid quark
stars. Even in case of solid quark stars, the GW amplitude is
relevant to the quadrupole deformations~\cite{Owen05} (e.g.,
mountain building on stellar surface) sustained by elastic or
magnetic forces on stellar surface. A quark or neutron star with
quadrupole deformation would be a GW radiator if it has precession
either free or torqued, and the precession amplitude (or the angle
between spin axis and spindle of inertia ellipsoid) is determined by
EoS~\cite{prec} and determines GW strain.
(ii). {\em GW being mass-dependent}. A very difference between quark
and normal neutron stars is that the latter is gravity-bound while
the former is confined additionally by self strong interaction, that
results in the fact that quark stars could be very low
massive~\cite{xu05} (even to be of $\sim 10^{-3}M_\odot$) but
neutron star cannot. Low mass quark stars, either in liquid or solid
states, are surely very weak GW emitters. This mass-depend nature
makes it more complex to constrain EoS of target compact stars by
negative results of LIGO GW detections.
The points of above (i) and (ii) are certainly very useful for us to
observationally distinguish quark stars from normal neutron stars in
the future.

Pulsars spin usually at frequencies $>10^0$ Hz, and we thus are
interested in LIGO to detect their GWs, from Fig.~\ref{fig:sens}.
There are two kinds of GWs from pulsar-like compact stars:
continuous GWs due to spin and bursting GWs due to stellar
catastrophic events (e.g., star quake~\cite{xty06} or binary
coalescence).
It is worth noting that all the upper limits estimated from LIGO
science runs depend on simulated waveform types of GWs (i.e.,
astrophysical GW radiative mechanisms).
For continuous GWs, the waveforms could be better understood, and
their searches are significantly more sensitive, especially when
informed by observational photon astronomy and theoretical
astrophysics~\cite{Owen09}.
The waveform of bursting GW is a matter of debate~\cite{Kalmus09},
and such kind of GW searches is also focused by LIGO, especially on
the super-flares of soft gamma-ray repeaters~\cite{Horvath05}.

\section{Summary and Future prospects}
\label{sec:fur}

Pulsars could be useful tools to detect GWs by PTA technique, they
would also be strong GW radiators; a real detection of gravitational
wave should be very meaningful in gravity physics, micro-theory of
elementary strong interaction, and astronomy.
A successful detection of GW by PTA may provide a test of GW
polarization and measure the graviton mass. Thought indirect
evidence for GWs from pulsar timing in binaries has been obtained, a
direct GW detection of pulsar-like stars is also expected as
persistent or transient sources. The strain of GW from an isolated
compact star depends on the equation of state of cold matter at
supra-nuclear densities.

Pulsar timing array projects are promising for detect GWs. We can
achieve $h_{c}=10^{-15}$ region for several year continuous pulsar
timing monitoring. If bi-weekly observations are made for five years
with RMS timing accuracy of 100 ns, then 40 pulsars are required for
general relativistic modes, 60 for the longitudinal mode; 60 for the
two spin-1 ¡°shear¡± modes; and 40 for the spin 0 ¡°breathing¡±
mode.
Additionally, we may measure the graviton mass through PTA
techniques. With a 5-year observation of 100 or 300 pulsars, we can
detect the graviton mass being higher than $2.5 \times 10^{-22}$ and
$10^{-22}$ eV, respectively. Ultimately, a 10-year observation of
300 pulsars allows us to probe the graviton mass at a level of $3
\times 10^{-23}$ eV.

For the task of measuring the GW polarization and the graviton mass,
there is one critical requirement: a large sample of stable pulsars.
Thus the on-going and coming projects like the Parkes
PTA~\cite{HBBC09}, the European PTA~\cite{SKLDJ06}, the Large
European Array for Pulsars~\cite{BVK09}, the FAST~\cite{NWZZJG04,
SLKMSJN09} and the SKA would offer unique opportunities to detect
the GW background and to probe into the nature of GWs, both physical
(the GW polarization and the graviton mass) and astronomy (the GW
sources).

Other important requirements for a successful PTA include high
stability of pulsar intrinsic noises and a low measurement noise. We
need pulsar survey with better sky coverage as well as good
observing system, especially with better band width\cite{YHC07} to
get better signal to noise ratio and to subtract the interstellar
medium effects. Better radio frequency interference filtering
technology will also be very helpful such that we can use the full
band data and reduce the terrestrial contamination. Better timing
techniques (such as timing in full Stokes parameters
\cite{VStraten06}) could also be preferred.

\section*{Acknowledgments}
We would like to acknowledge useful discussions at the pulsar group
of PKU.
This work is supported by NSFC (10833003, 10973002), the National
Basic Research Program of China (grant 2009CB824800) and LCWR
(LHXZ200602).


\end{document}